\documentclass[
reprint, 
superscriptaddress, 
amsmath, 
amssymb, 
aps, 
longbibliography]
{revtex4-2}

\usepackage{graphicx}
\usepackage{caption}
\usepackage{dcolumn}
\usepackage{bm}
\usepackage{gensymb}
\usepackage{physics}
\usepackage[letterpaper, margin=0.6in]{geometry}
\usepackage{xcolor}
\usepackage{amsmath}

\usepackage{hyperref}
\usepackage[utf8]{inputenc}

\renewcommand{\rmdefault}{ptm} 

\hypersetup{
    colorlinks,
    citecolor=black,
    filecolor=black,
    linkcolor=black,
    urlcolor=black
}

\begin{document}
\captionsetup[figure]{labelfont={normal},labelformat={default},labelsep=period,name={FIG.},justification=raggedright,singlelinecheck=false}

\title{Superconducting Dome in Ionic Liquid Gated Homoepitaxial Strontium Titanate Thin Films}

\author{Sushant Padhye}
\affiliation{Department of Electrical and Computer Engineering, University of Cincinnati, Cincinnati, OH 45219, USA}

\author{Jin Yue}
\affiliation{Department of Chemical Engineering and Materials Science, University of Minnesota, Minneapolis, MN 55455, USA}

\author{Shivasheesh Varshney}
\affiliation{Department of Chemical Engineering and Materials Science, University of Minnesota, Minneapolis, MN 55455, USA}

\author{Bharat Jalan}
\affiliation{Department of Chemical Engineering and Materials Science, University of Minnesota, Minneapolis, MN 55455, USA}

\author{David Goldhaber-Gordon}
\affiliation{Department of Physics, Stanford University, Stanford, CA 94305, USA}
\affiliation{Stanford Institute for Materials and Energy Sciences, SLAC National Accelerator Laboratory, Menlo Park, CA 94025, USA}

\author{Evgeny Mikheev}
\affiliation{Department of Physics, University of Cincinnati, Cincinnati, OH 45221, USA}
\affiliation{Department of Physics, Stanford University, Stanford, CA 94305, USA}
\affiliation{Stanford Institute for Materials and Energy Sciences, SLAC National Accelerator Laboratory, Menlo Park, CA 94025, USA}

\begin{abstract}

In this work, we patterned a two-dimensional electron gas (2DEG) on the surface of a SrTiO$_3$ thin film grown homoepitaxially on SrTiO$_3$ by hybrid molecular beam epitaxy (hMBE). We explored the superconducting dome in this material system by tuning electron density with ionic liquid gating. We found superconducting transitions up to 503 mK near an optimal electron density of approximately 3 $\times$  10$^{13}$ cm$^{-2}$. This is a meaningful increase from the typical optimal transition near 350 mK in similar 2DEGs on SrTiO$_3$ single crystal substrate surfaces. Systematic tuning of 2DEG electron density revealed a consistent BCS scaling between superconducting critical temperature, coherence length, and electron mean free path. Substantial variation of transition width across the dome was described by a paraconductivity model combining Aslamazov-Larkin and Maki-Thompson contributions.

\noindent \textbf{Corresponding author:} Evgeny Mikheev, Email: mikheev@ucmail.uc.edu.

\end{abstract}
\maketitle

\medskip


\section{Introduction}

The superconducting dome of SrTiO$_3$  hosts an interwoven web of intriguing material physics. Modern systematic studies in uniformly doped crystals and thin films revealed complex interplay of superconductivity with material structure \cite{19_behnia_annrev_cmp}, and superconducting transition point enhancement by proximity to ferroelectricity \cite{19_stemmer_sciadv, 22_hameed_natmatlet, 19_tomioka_ncomms, 19_herrera_physrevmat}. The dome extends into an unusually dilute carrier density regime, a feature that was identified as challenging to reconcile with the BCS framework \cite{13_behnia_physrevx, 20_gastiasoro_review_elsevier}. The dilute nature of SrTiO$_3$ superconductivity also allows for electrostatic tuning of superconducting 2D electron gas systems (2DEG). This has been extensively explored for interface 2DEGs in SrTiO$_3$/LaAlO$_3$ and related heterostructures, primarily by back-gating through the SrTiO$_3$ substrate \cite{08_caviglia_natlett, 22_bergeal_advmatint, 12_bergeal_physrevlet, 12_joshua_natcom}. The dominant effect on superconductivity in such a setup typically arises from confinement profile modulation, rather than electron density tuning \cite{16_chen_acsnanolett}.

In this work, a 2DEG is accumulated on the surface of an undoped SrTiO$_3$ thin film by ionic liquid gating. This technique \cite{21_mikheev_sciadv, 22_mikheev_nanolett, 23_mikheev_natelec} enables systematic tuning of the electron density across the superconducting dome in a single Hall bar device and superconducting transition characterization in temperature and magnetic field. The SrTiO$_3$ thin film was grown by hybrid molecular beam epitaxy, a deposition method that allows for precise control over film stoichiometry and minimizes structural defects levels \cite{09_jalan_apphyslet, 09_jalan_vacscitech, 10_jalan_natmatlet, 22_jalan_sciadv}.

In uniformly doped SrTiO$_3$ crystals and thin films, the superconducting transition ($T_\mathrm{c}$) at the peak of the superconducting dome is 500 mK or less \cite{14_behnia_physrevlett, 15_behnia_physrevb}. Recently, it was discovered that $T_\mathrm{c}$ can be enhanced by bringing SrTiO$_3$, an incipient ferroelectric, closer to ferroelectricity, suggesting a possible ferroelectric quantum criticality scenario \cite{15_edge_physrevlett, 20_gastiasoro_review_elsevier}. $T_{\mathrm{c}}$ enhancement was observed up to 660 mK with compressive epitaxial strain \cite{19_stemmer_sciadv}, 600 mK by uniaxial strain \cite{19_herrera_physrevmat}, 620 mK by plastic deformation \cite{22_hameed_natmatlet}, and 550-600 mK by Ca and oxygen isotope substitution \cite{19_tomioka_ncomms, 22_rischau_physrevres}. In SrTiO$_3$ surface and heterostructure 2DEG systems, the optimal $T_\mathrm{c}$ is typically near 350 mK \cite{12_joshua_natcom}, and no conclusive $T_\mathrm{c}$  enhancement by ferroelectricity has been demonstrated so far. In comparison to a maximum $T_\mathrm{c}$ of 370 mK in similar ionic liquid gated 2DEG devices on SrTiO$_3$ single crystal surfaces \cite{21_mikheev_sciadv}, we observe here an unexpected increase of $T_\mathrm{c}$ to 500 mK in a 2DEG on the surface of a homoepitaxial SrTiO$_3$ film.

Despite the unconventionally dilute, multiband, and quantum critical nature of SrTiO$_3$ superconductivity, some of its features appear remarkably conventional. The magnitude and temperature dependence of its superconducting gap aligns  closely with the conventional BCS description in the weak coupling limit \cite{18_swartz_pnas}. In this work, we further reinforce the evidence for a BCS description of superconductivity in SrTiO$_3$, by mapping electron density dependence of the superconducting coherence length scale, and demonstrating that it follows BCS scaling across the entire superconducting dome. Systematic transport measurements also enabled us to formulate a unified description for excess conductivity above $T_{\mathrm{c}}$ from conventional superconducting fluctuations.

\section{Methods}

\begin{figure}
\centering
\includegraphics[width=3.5in]{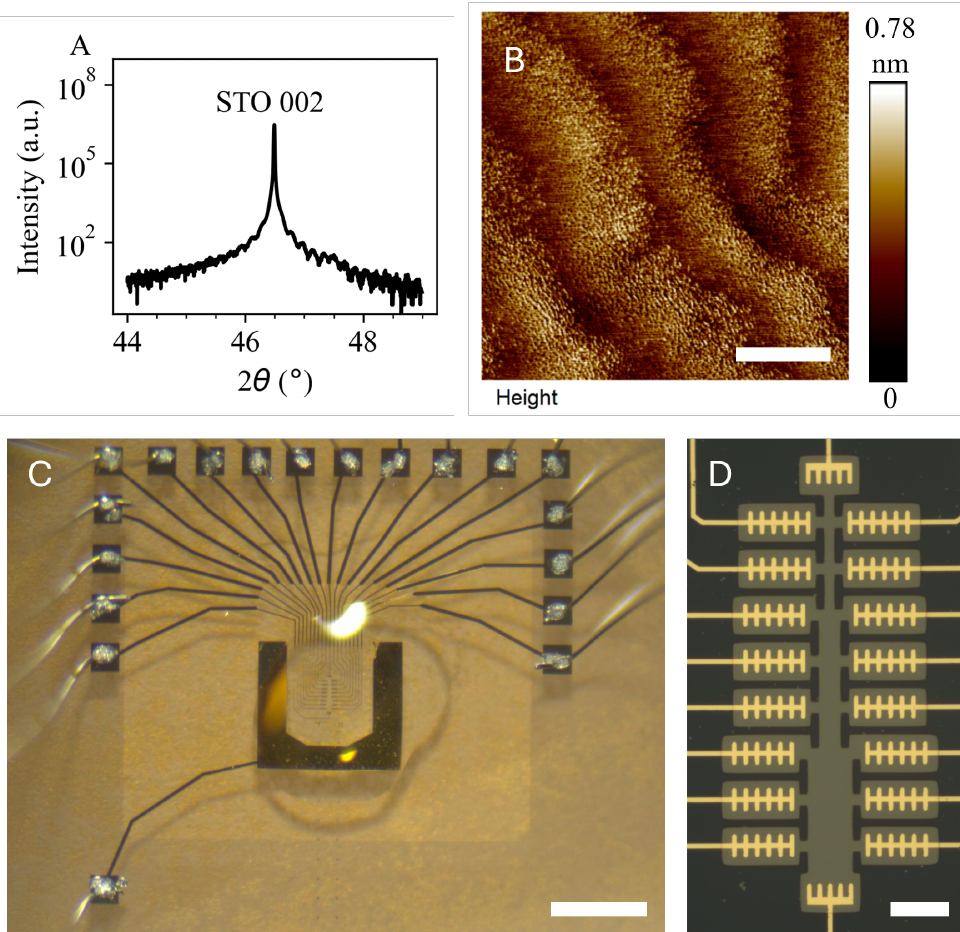}
\caption{\label{growth_figure} \textbf{Film growth and ionic liquid gated device.} (A) High-resolution x-ray diffraction (XRD) 2$\theta$-$\omega$ coupled scan, overlapped SrTiO$_3$ substrate and film peaks. (B) Atomic Force Microscopy (AFM) image of the sample after growth. (C) Picture of the device with ionic liquid deposited on the device. (D) Picture of the fabricated Hall Bar device. The dark, light and golden ($\textcolor[HTML]{31332e}{\blacksquare}$, $\textcolor[HTML]{67644b}{\blacksquare}$, $\textcolor[HTML]{e5ca72}{\blacksquare}$) areas correspond to the insulating SiO$_2$, exposed STO, and Au/Ti contacts, respectively. 
\newline
Scale bars in (B), (C), (D) are 1 $\mu$m, 1 mm and 50 $\mu$m, respectively.}
\end{figure}

A 60-nm thick, undoped, and electrically insulating SrTiO$_3$ thin film was grown on an undoped STO (001) substrate by hybrid molecular Beam Epitaxy (hMBE) involving a metal-organic precursor as a source of Ti. This technique allows for synthesis of very high quality thin films by tuning Sr and Ti within the ``growth window" with self-regulating stoichiometry. The overlap of the SrTiO$_3$ substrate and film peak in the x-ray diffraction scan in Figure \ref{growth_figure}A indicates a near perfect stoichiometry, as any Ti/Sr or oxygen non-stoichiometry expands the SrTiO$_3$ unit cell \cite{09_jalan_applphyslett}. The film surface is atomically smooth on an atomic force microscope (AFM) image in Figure \ref{growth_figure}B. 

The lithographically defined Hall bar device shown in Figure \ref{growth_figure}C consists of metallic ohmic contact lines, a large U-shaped gate contact, and a patterned insulating SiO$_2$ layer. The latter defines the contour of the 2DEG channel that is formed by ionic liquid gating on exposed SrTiO$_3$ surfaces, while areas covered by SiO$_2$ remain electrically insulating (Fig. \ref{growth_figure}D). Hall bar channel section with different widths of 10, 20 and 40 $\mu$m were defined to confirm 2DEG homogeneity. As described in supplementary section S3, electron mobility and Hall density above $2 \times 10^{13}$ cm$^{-2}$ were observed to be uniform along the channel

Following the techniques developed in \cite{21_mikheev_sciadv, 23_mikheev_natelec}, both the U-shaped gate and the Hall bar are covered by an ionic liquid DEME-TFSI. The voltage on the ionic liquid gate contact  $V_{\mathrm{GIL}}$ was ramped under vacuum at room temperature inside a sample insert of a dilution refrigerator. A gradual transition from an insulator to a conductor started around $V_{\mathrm{GIL}}$ = 1 V. During the subsequent cooldown process, the sample was maintained at a constant $V_{\mathrm{GIL}}$. The ionic liquid freezes near 200 K, fixing the 2DEG charge until the sample is thermally cycled back to room temperature to adjust $V_{\mathrm{GIL}}$. We performed five cooldowns with different $V_{\mathrm{GIL}}$ set points to study the superconducting dome. Measurements were performed at low frequencies (15 to 30 Hz) and low AC and DC excitation currents (10 nA and up to 2 $\mu$A respectively). After cooling down to a base temperature near 10 mK, $V_{\mathrm{GIL}}$ was ramped up to 10 V, resulting in ohmic contact resistance reduction and marginal electron mobility increase via a back gate-like effect (see \cite{23_mikheev_natelec} and supplementary information S4 B). 

\section{Results and Discussion}

\subsection{Enhancement of superconductivity}

\begin{figure*}[htbp]
\centering
\includegraphics[width=7in]{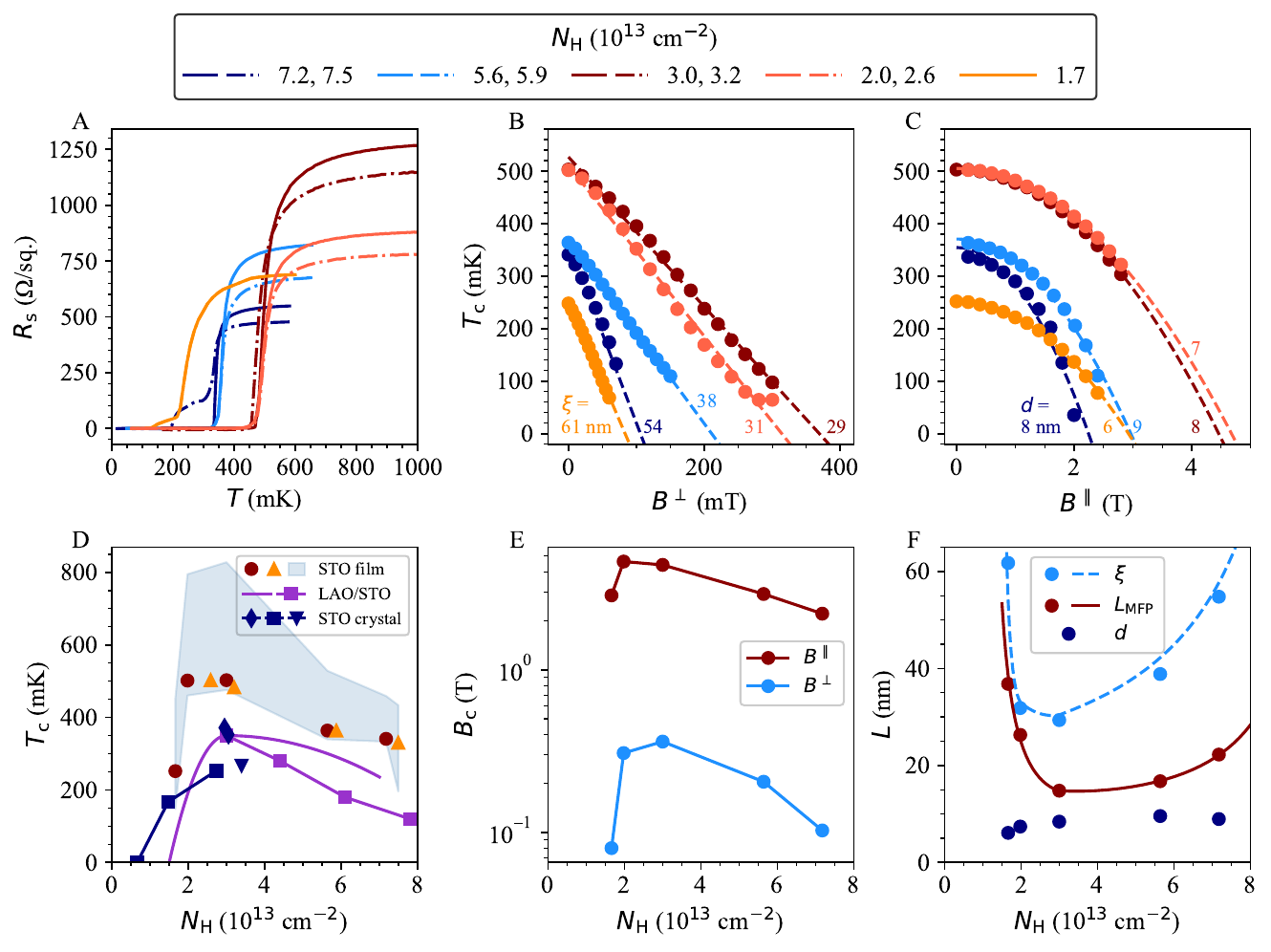}
\caption{\label{panel_tcnh_gltheory} \textbf{Superconducting dome in ionic liquid gated homoepitaxial SrTiO$_3$ films.} (A) Superconducting transitions with temperature. Solid and dot-dashed lines are for the 10 and 40 $\mathrm{\mu}$m channels respectively. For the lowest carrier density, measurements were across the entire hall bar with the 10, 20, and 40 $\mu$m channel width segments in series. (B), (C) Critical temperature as a function of magnetic field shown along with dashed line fits to G-L theory using Equations \ref{eqbperp} and \ref{eqbpar}. $\xi$ and $d$ fit values are also indicated for each fit. (D) Critical temperatures extracted as the point at which $R_{\mathrm{s}} = 0.5\times R_{\mathrm{n}}$ ($R_{\mathrm{n}}$ is the normal-state resistance). Shaded region shows the width of the transitions as $T(R_{\mathrm{s}} = 0.9\times R_{\mathrm{n}})-T(R_{\mathrm{s}} = 0.1\times R_{\mathrm{n}})$. $\textcolor[HTML]{8b0000}{\bullet}$ and $\textcolor[HTML]{ff8d02}{\blacktriangle}$ show the critical temperature for the 10 and 40 $\mathrm{\mu}$m channels respectively. The blue line is a guide to the eye. LAO/STO data are from \cite{10_shalom_physrevlet}. The LAO/STO dome interpolation follows \cite{22_bergeal_advmatint, 23_mikheev_natelec}. Data for ionic liquid gated devices on STO single crystal surfaces are from \cite{21_mikheev_sciadv}. (E) Critical magnetic field ($B_{\mathrm{c}}$) extrapolated from the fits shown as a function of $N_{\mathrm{H}}$. (F) $\xi$, $L_{\mathrm{MFP}}$ and $d$ dependence with $N_{\mathrm{H}}$. $\xi$ was obtained from G-L framework and through independent measurement and calculations based on BCS theory (shown with the dashed line using interpolated points).}
\end{figure*}

\begin{figure*}[htbp]
\centering
\includegraphics[width=7in]{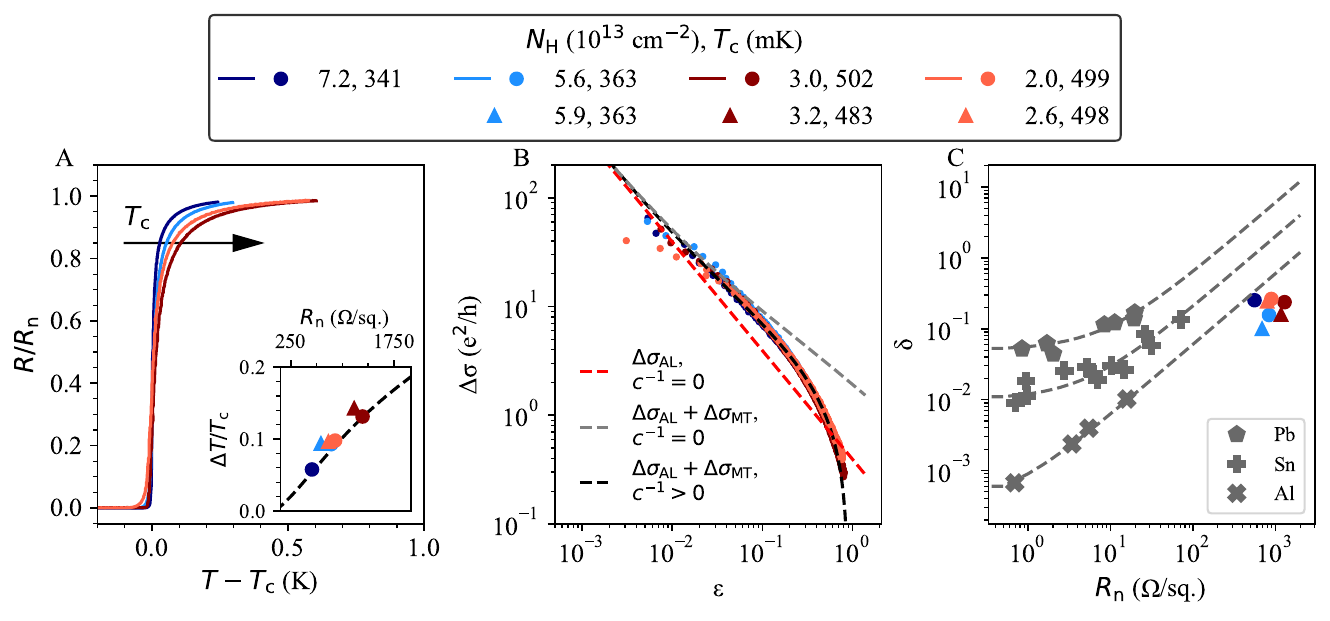}
\caption{\label{figure_al_mt_fits} \textbf{Paraconductivity above the superconducting transition.} (A) Normalized transitions are shown for the 10 $\mu$m channel. The arrow signifies that the broader transitions have a higher $T_{\mathrm{c}}$. In subsequent plots, the $\bullet$ and $\blacktriangle$ markers show the 10 and 40 $\mu$m channels respectively. The inset shows the width of the transition taken as $\Delta T = T(R = 0.9*R_{\mathrm{n}}) - T_{\mathrm{c}}$. The dashed line is the expected width from AL-MT fluctuations ($\delta=0.22$ and $c=1$) while varying only $R_{\mathrm{n}}$. (B) Superconducting transitions shown as change in conductivity with $\varepsilon=ln(T/T_{\mathrm{c}})$. Dashed lines show the AL contribution only (red), AL and MT contributions combined using  $\delta=0.22$ (grey), same AL and MT contributions with a total-energy cutoff parametrized by $c=1$ (black). (C) The dependence of $\delta$ with $R_{\mathrm{n}}$ is shown. The $\textcolor[HTML]{696969}{\bullet}, \textcolor[HTML]{696969}{\blacksquare}, \textcolor[HTML]{696969}{\blacktriangle}$ markers show the measured data along with dashed lines as linear fits for other superconductors \cite{72_crow_physrevlett}.}
\end{figure*}

We observed superconducting transitions for all cooldowns as shown in Figure \ref{panel_tcnh_gltheory}D. Notably, with $T_{\mathrm{c}}$ defined as the midway point of the resistive transition from normal state resistance to zero, its optimal value near 2-3 $\times$ 10$^{13}$ cm$^{-2}$ was slightly above 500 mK (503 mK at 2.6 $\times$ 10$^{13}$ cm$^{-2}$). The Hall densities ($N_\mathrm{H}$) at which superconductivity onsets and at which it reaches optimal $T_{\mathrm{c}}$ are consistent with previous reports on SrTiO$_3$ heterostructures and ionic liquid 2DEGs \cite{12_joshua_natcom, 21_mikheev_sciadv}. But the optimal $T_\mathrm{c}$ value shows a significant increase in comparison to other SrTiO$_3$ 2DEG systems (Figure \ref{panel_tcnh_gltheory}D). In particular, very similar ionic liquid gated 2DEG devices on STO single crystal surfaces consistently showed $T_\mathrm{c}$ lower than 370 mK \cite{21_mikheev_sciadv, 22_mikheev_nanolett}. $T_\mathrm{c}$ enhancement by 100-200 mK in comparison to STO single crystal devices and SrTiO$_3$/LaAlO$_3$ is observed over a broad electron density range that includes the dome peak and the ``overdoped" regime. Similarly to other SrTiO$_3$ 2DEG reports \cite{11_caprara_physrevb, 19_caprara_physrevb}, these superconducting transitions showed substantial widths in temperature (shaded grey area in Figure \ref{panel_tcnh_gltheory}A), but with a pronounced asymmetry and a gradual onset above the extracted $T_\mathrm{c}$ point. As discussed in further detail below (see section III-C), we attribute transition broadening above $T_\mathrm{c}$ to paraconductivity from superconducting fluctuations, rather than non-uniformly enhanced $T_\mathrm{c}$.

We attribute the observed $T_\mathrm{c}$ enhancement to the only deliberate change in our study with respect to single crystal STO device work in \cite{21_mikheev_sciadv}, the addition of the homoepitaxial STO thin film. SrTiO$_3$ thin films grown by hMBE within the self-regulating stoichiometry window contain a reduced amount of unintentional defects and impurities such as Al and Fe in comparison to commercial single crystal substrates \cite{10_jalan_natmatlet}. Although disorder reduction is a possible factor in the observed $T_\mathrm{c}$ enhancement, studies using uniformly doped homoepitaxial SrTiO$_3$ thin films \cite{22_jalan_sciadv}, and electron irradiation in single crystal found that disorder does not have an observable effect on $T_\mathrm{c}$, consistent with s-wave superconductivity \cite{15_behnia_physrevb}.

A related and possibly important factor is compressive microstrain caused by minuscule differences in lattice parameter from disorder reduction in the SrTiO$_3$ film versus the substrate. $T_\mathrm{c}$ enhancement up to 660 mK was observed in compressively strained SrTiO$_3$ films with uniform doping \cite{19_stemmer_sciadv} and upon compressive plastic deformation \cite{22_hameed_natmatlet}, consistent with superconductivity enhancement by bringing SrTiO$_3$ closer to ferroelectric quantum criticality \cite{15_edge_physrevlett, 20_gastiasoro_review_elsevier}. The same $T_\mathrm{c}$ enhancement mechanism could be at play in our thin film surface 2DEG device. Alternatively, a small microstrain is likely to affect the spontaneous formation of the tetragonal domain wall network below the cubic-to-tetragonal transition near 100 K \cite{22_jalan_sciadv}. Inhomogeneity of normal state and superconducting transport linked to such domain walls has been extensively documented in STO \cite{13_moler_natmatlett, 16_moler_physrevb, 17_kalisky_natmat}, suggesting a plausible coupling mediator between microstrain and superconductivity.

\subsection{Superconducting Length Scale Evolution}

In addition to characterizing the temperature driven superconducting transitions, we also studied the superconducting coherence length and 2DEG thickness and their relation with carrier density. Measurements were performed with magnetic field in both the out-of-plane ($B^\perp$) and in-plane ($B^\parallel$) axes. Holding magnetic field constant, we measured sheet resistance $R_{\mathrm{s}}$ as a function of temperature $T$. This allowed us to track $T_{\mathrm{c}}$ as a function of magnetic field for our cooldowns and carry out extrapolations to $B=0$ using the Ginzburg-Landau framework for a 2D superconducting layer \cite{12_hwang_physrevb, 17_monteiro_physrevb}:

\begin{equation}
\label{eqbperp}
T_{\mathrm{c}} = \left(1 - B^\perp \cdot \frac{2\pi\xi^2}{\phi_{\mathrm{0}}}\right) \cdot T_{\mathrm{c}}(B=0),
\end{equation}
and 
\begin{equation}
\label{eqbpar}
T_{\mathrm{c}} = \left[1 - \left(B^{\parallel} \cdot \frac{2\pi\xi d}{\sqrt{12}\phi_{\mathrm{0}}}\right)^2\right] \cdot T_{\mathrm{c}}(B=0).
\end{equation}

$\phi_{\mathrm{0}}$ is the flux quantum, $\xi$ is the superconducting coherence length, and $d$ is the superconducting layer thickness. The fits to $B^\perp$ dependence of $T_{\mathrm{c}}$ to  Equation \ref{eqbperp} gives $\xi$ (Figure \ref{panel_tcnh_gltheory}B).
The subsequent fit of $B^\parallel$ dependence of $T_{\mathrm{c}}$ to Equation \ref{eqbpar} gives $d$ (Figure \ref{panel_tcnh_gltheory}C). 

The fitted values for $d$ are 6-9 nm and do not substantially change with $N_{\mathrm{H}}$, pointing to systematic tuning of $N_{\mathrm{H}}$ without significant changes to the confinement profile. This is an advantage for the ionic liquid gating technique used here for systematic study of $N_{\mathrm{H}}$ dependence over back-gating through the SrTiO$_3$ substrates, which typically results in predominant tuning of the 2DEG thickness rather than Hall density \cite{16_chen_acsnanolett, 18_hwang_ncomms}. The $d$ in our device is also consistently  smaller than $\xi$ of 30 nm or above, as shown in Figure \ref{panel_tcnh_gltheory}F, confirming that this is a two dimensional superconducting system.

Zero temperature extrapolations of the critical fields ($B_{\mathrm{c}}$) from fits to equations \ref{eqbperp} and \ref{eqbpar} are shown in Figure \ref{panel_tcnh_gltheory}E. Maximizing in-plane critical fields is technologically important for oxide Majorana device proposals \cite{18_caprara_mdpi, 19_perroni_physrevb}. Near the peak of the superconducting dome, the extrapolated in-plane $B_{\mathrm{c}}$ is approximately 4.3 T (above our maximum experimentally available 3 T field in the horizontal direction) and the form of equation \ref{eqbpar} suggests that further enhancement is possible by heterostructure engineering narrower 2DEG confinement \cite{11_stemmer_applphyslet, 12_stemmer_physrevb, 13_stemmer_applyphyslet}.

Previous tunneling studies \cite{18_swartz_pnas} observed that SrTiO$_3$ is accurately described by the weak coupling limit of BCS theory with a single superconducting gap $\Delta_0 = 1.764k_{\mathrm{b}}T_{\mathrm{c}}$. In this limit, the intrinsic superconducting coherence length scale is:

\begin{equation}
\label{eqxiBCS}
\xi_{\mathrm{BCS}} = \hbar v_{\mathrm{F}}/\pi\Delta_0,
\end{equation}

where the two-dimensional Fermi velocity is $v_{\mathrm{F}} = \hbar k_{\mathrm{F}}/m^*$. SrTiO$_3$ is typically in the ``dirty" limit \cite{15_behnia_physrevb}, where the disorder length  scale is smaller than $\xi_{\mathrm{BCS}}$, the effective coherence length scale becomes:

\begin{equation}
\label{eqxi}
1/\xi = \frac{2\sqrt{3}}{\pi}\frac{\sqrt{1+\xi_{\mathrm{BCS}}/L_{\mathrm{MFP}}}}{\xi_{\mathrm{BCS}}}
\end{equation}

The electron mean free path $L_{\mathrm{MFP}}$ can be estimated from $v_{\mathrm{F}}$ and the time between scattering: $L_{\mathrm{MFP}} = v_{\mathrm{F}}\tau = \mathrm{\mu}_{\mathrm{H}}e^{-1}\hbar\sqrt{2\pi N_{\mathrm{H}}}$. This calculated mean free path is shown in Figure \ref{panel_tcnh_gltheory}F with a line for polynomial interpolation between data points. Hall electron mobility $\mathrm{\mu}_{\mathrm{H}}$ ranged between 150 and 550 cm$^2$/V$\cdot$s in our devices, corresponding to $L_{\mathrm{MFP}}$ between 15 and 40 nm. This is comparable with other 2DEGs in STO heterostructures and ionic liquid gated STO single crystal surfaces \cite{23_mikheev_natelec, 18_thierschmann_ncomms, 20_bergeal_natelec, 20_jespersen_physrevmat}.

The dashed line in Figure \ref{panel_tcnh_gltheory}F is a weak coupling BCS limit prediction for the effective coherence length $\xi$ from Equations \ref{eqxiBCS} and \ref{eqxi}. It uses as input experimentally measured $B^\perp,T_{\mathrm{c}},N_{\mathrm{H}},L_{\mathrm{MFP}}$. The only adjustable parameter is the effective mass $m^* = 5m_{\mathrm{e}}$. The observed good agreement between calculated and measured $\xi$ in our experiment supports applicability of the BCS description for SrTiO$_3$ 2DEGs. But the success of fixing a single value of effective mass across the entire range of $N_{\mathrm{H}}$ in our experiment is surprising: a Lifshitz transition is typically observed in STO 2DEGs around 2-3 $\times$ 10$^{13}$ cm$^{-2}$, from single light $d_{\mathrm{xy}}$ band to multi-band occupation with heavier $d_{\mathrm{xz/yz}}$ electrons \cite{10_lee_physrevb, 12_bergeal_physrevlet, 22_bergeal_advmatint}. A single $m^* = 5m_{\mathrm{e}}$ value implies heavy band occupation across the entire superconducting dome. Such band order inversion has been observed in high mobility $\gamma$-Al$_2$O$_3$/STO 2DEG heterostructures \cite{21_chikina_nanolett}. Multi-band occupancy in STO 2DEGs is commonly associated with non-linearity in Hall effect \cite{10_lee_physrevb, 12_bergeal_physrevlet}. However, no significant deviation from linear Hall effect was observed in our case up to $B^\perp$ = 9 T ($<$15\% at the lowest carrier density and $<$30\% at the highest), consistent with single heavy band occupation.

\subsection{Superconducting Fluctuations above the Transition}

We observed that the high $T_{\mathrm{c}}$ cooldowns near the peak of the superconducting dome appeared to have broader superconducting transitions. Normalization in Figure \ref{figure_al_mt_fits}A highlights that these transitions have a short ``tail" below $T_{\mathrm{c}}$ and a long ``head" - a gradual onset of the resistance drop well above $T_{\mathrm{c}}$. The inset of Figure \ref{figure_al_mt_fits}A shows that the width of this transition head (defined as temperature difference $\Delta T =$ $T(R = 0.9*R_{\mathrm{n}}) - T_{\mathrm{c}}$)  scales with the normal state resistance. The analysis below describes the trend line in this inset in terms of combined Aslamazov-Larkin (AL) and Maki-Thompson (MT) contributions to thermodynamic fluctuations of the superconducting order parameter \cite{75_tinkham_iop}.

This framework is well established for conventional metallic superconductors in reduced dimensions, giving excess conductivity (paraconductivity) $\Delta\sigma=R^{-1}-R_n^{-1}$ above $T_{\mathrm{c}}$. Following \cite{19_mosqueira_physrevb}, our analysis uses two-dimensional AL and MT expressions with $\Delta\sigma=\Delta\sigma_{\mathrm{AL}}+\Delta\sigma_{\mathrm{MT}}$ with reduced temperature $\varepsilon = ln(T/T_{\mathrm{c}})$.

Importantly, re-plotting experimental $R(T)$ data above $T_{\mathrm{c}}$ from cooldowns across the superconducting dome as $\Delta\sigma(\varepsilon)$ collapses them onto a single curve (Figure \ref{figure_al_mt_fits}B), strongly suggesting a unified paraconductivity mechanism. We note that no fine-tuning was necessary to produce this collapse. It is robust in  the intermediate range ($10^{-2}>\varepsilon>10^{-1}$), but very sensitive to the extraction procedure for $T_{\mathrm{c}}$ and $R_{\mathrm{n}}$ values at small and large $\varepsilon$, respectively. Data collapse in \ref{figure_al_mt_fits}B was produced by systematically using the peak of $dR/dT$ for $T_{\mathrm{c}}$ and  the saturation values of $R_{\mathrm{s}}$ at  high $T$ and $B^\perp > B_{\mathrm{c}}$ for $R_{\mathrm{n}}$. 

The ``direct" AL contribution comes from the direct acceleration of Cooper pairs created by fluctuations above $T_{\mathrm{c}}$ and follows a universal curve that is only scaled by $T_{\mathrm{c}}$:
\begin{equation}
\label{eq_al_c}
\Delta\sigma_{\mathrm{AL}}(\varepsilon) = \frac{e^2}{16\hbar}\cdot\left(\frac{1}{\varepsilon}-\frac{1}{c}\right).
\end{equation}
In both AL and MT expressions, we included the total energy cutoff parameter $c$ which discards paraconductivity from short wavelength fluctuations far above $T_{\mathrm{c}}$ \cite{19_mosqueira_physrevb}. It also accounts for uncertainty in determination of $R_{\mathrm{n}}$. The obtained values for $c$ from fits are shown in Figure S11.

The ``indirect" MT contribution is associated with the decay of fluctuating Cooper pairs into quasiparticles. Its magnitude is set by $\delta = (T_{\mathrm{c0}}-T_{\mathrm{c}})/T_{\mathrm{c}}$, a parametrization of a transition temperature shift from pair-breaking interactions:
\begin{equation}
\label{eq_mt_c}
\Delta\sigma_{\mathrm{MT}}(\varepsilon) = \frac{e^2}{8\hbar}\cdot\left[\frac{ln(\varepsilon/\delta)}{\varepsilon-\delta}-\frac{ln(c/\delta)}{c-\delta}\right].
\end{equation}

The red dashed line in Figure \ref{figure_al_mt_fits}B is the universal AL curve without the cutoff ($c^{-1}=0$), showing that the AL model alone fails to account for the entirety of excess conductivity. The black dashed line is an illustrative fit line for $\Delta\sigma_{\mathrm{AL}}+\Delta\sigma_{\mathrm{MT}}$ using $\delta = 0.22$ and $c = 1$ as fitting parameters for the cooldown at $N_{\mathrm{H}}=3\times10^{13}\mathrm{cm}^{-2}$, showing good agreement with experimental data for all cooldowns. The gray dashed line is a $\Delta\sigma_{\mathrm{AL}}+\Delta\sigma_{\mathrm{MT}}$ curve using the same $\delta$ but without the cutoff. This trace illustrates that the AL+MT curve with $\delta$ as the only fit parameter matches the magnitude and slope of $\Delta\sigma$ below $\varepsilon\approx 0.1$.

Due to a weak dependence of $\delta$ on equation \ref{eq_mt_c}, even small differences between experimental $\Delta\sigma(\varepsilon)$ traces in Figure \ref{figure_al_mt_fits}B produce a significant scatter between 0.1-0.3 in the obtained $\delta$ values. They are plotted in Figure \ref{figure_al_mt_fits}C against normal state resistance $R_{\mathrm{n}}$ for comparison with conventional superconductors. This figure also shows data for Pb, Sn, and Al along with a linear trend line from \cite{72_crow_physrevlett}. This trend is understood as a combination of a pair breaking term that linearly scales with $R_n$ and a constant ``thermal" contribution $\delta_\mathrm{th}$ that varies across materials, but for a given material is independent of $R_\mathrm{n}$ \cite{72_crow_physrevlett, 75_tinkham_iop}. Empirically, both terms increase with $T_{\mathrm{c}}$ and our SrTiO$_3$ 2DEG data falls within this trend, slightly below Al $\delta(R_{\mathrm{n}})$ curve. But since SrTiO$_3$ 2DEGs are superconductors with unusually low electron density, this comparison requires a wide extrapolation of Pb, Sn, and Al trend lines in $R_{\mathrm{n}}$ above the limit of experimental validation \cite{72_crow_physrevlett, 19_mosqueira_physrevb}. Our study is limited to a narrow range of $R_{\mathrm{n}}$, and a complete understanding of the MT contribution in SrTiO$_3$ will require a study of paraconductivity in superconducting 2DEGs with electron mobility higher than the measured values in our device \cite{15_gallagher_ncomms, 25_olsen_physrevmat}.

We also evaluated effective medium theory (EMT, \cite{11_caprara_physrevb, 19_caprara_physrevb}) as an alternate model for superconducting transition broadening with a bimodal $T_{\mathrm{c}}$ distribution. As described in section S6 of supplementary information, we conclude that Al-MT is the more likely explanation for the substantial asymmetric broadening above $T_{\mathrm{c}}$.
\section{Conclusions}

We demonstrated 2D electron density tuning by ionic liquid gating from the ``underdoped" to the ``overdoped" side of the superconducting dome. We found systematic agreement between measured coherence length scales and the BCS prediction. We discovered that transition broadening above $T_\mathrm{c}$ systematically collapses onto a universal paraconductivity curve following Aslamazov-Larkin and Maki-Thompson superconducting fluctuation theory. These aspects of robust conventional superconductor behavior add to the dichotomy with the unconventional dilute \cite{19_behnia_annrev_cmp} and quantum critical \cite{15_edge_physrevlett, 20_gastiasoro_review_elsevier} nature of SrTiO$_3$ superconductivity.

This work demonstrates viability of combining SrTiO$_3$ thin film growth by hybrid MBE with ionic liquid gating for creating highly tunable superconducting 2DEGs. This offers a blueprint for exploring correlation between SrTiO$_3$ film structure tailoring by MBE (e.g. with strain engineering, precise off-stoichiometry, A or B site co-doping, confinement profile tailoring through heterostructuring) and the superconducting dome. The observed optimal $T_{\mathrm{c}}$ enhancement from 350 mK (on single crystal surfaces) to 503 mK (on homoepitaxial film surface) highlights its high sensitivity to small changes in material structure. This emphasizes the need for precise control over SrTiO$_3$ surface and/or interface structure in future studies of 2DEG superconductivity.

\section{Acknowledgments}

We thank Yashar Komijani for insightful discussions and Dickson Boahen for assistance with cryogenic measurements. Cryogenic electrical measurements and data analysis at University of Cincinnati by S.P. and E.M. were supported by the office of Naval Research through award N00014-24-1-2079. Fabrication and preliminary cryogenic electrical measurements by E.M at Stanford University were supported by the U.S. Department of Energy, Office of Science, Basic Energy Sciences, Materials Sciences and Engineering Division, under Contract DE-AC02-76SF00515. Fabrication work was performed at the Stanford Nano Shared Facilities (SNSF)/Stanford Nanofabrication Facility (SNF), supported by the National Science Foundation under award ECCS-1542152. 

Growth and characterization of the film was primarily supported by the U.S. Department of Energy through grant nos. DE-SC0020211, and partly the Center for Programmable Energy Catalysis, an Energy Frontier Research Center funded by the U.S. Department of Energy, Office of Science, Basic Energy Sciences at the University of Minnesota, under Award No. DE-SC0023464. J.Y. acknowledges support from the Air Force Office of Scientific Research (AFOSR) through Grant No. FA9550-21-1-0025. Film growth was performed using instrumentation funded by AFOSR DURIP awards FA9550-18-1-0294 and FA9550-23-1-0085. Parts of this work were carried out at the Characterization Facility, University of Minnesota, which receives partial support from the NSF through the MRSEC program under Award No. DMR-2011401.

\section{Data availability}

Raw data, python notebooks used during original data collection, initial analysis, and generation of manuscript figures are available at https://doi.org/10.5281/zenodo.19264456.





\renewcommand{\rmdefault}{ptm} 

\hypersetup{
    colorlinks,
    citecolor=black,
    filecolor=black,
    linkcolor=black,
    urlcolor=black
}


\onecolumngrid
\captionsetup[figure]{labelfont={bf},labelformat={default},labelsep=period,name={Fig.}}
\setcounter{figure}{0}
\renewcommand{\thefigure}{S\arabic{figure}}
\setcounter{page}{1}
\renewcommand{\thepage}{S\arabic{page}}
\setcounter{section}{0}
\renewcommand{\thesection}{S\arabic{section}}
\setcounter{equation}{0}
\renewcommand{\theequation}{S\arabic{equation}}



\clearpage

\section*{Supplementary material for ‘‘Superconducting Dome in Ionic Liquid Gated Homoepitaxial Strontium Titanate Thin Films’’}

\section{Film Growth}

Epitaxial, single crystalline SrTiO$_3$ (001) films were grown with a hybrid MBE (hMBE) approach \cite{22_jalan_sciadv}. The 5 mm × 5 mm × 0.5 mm SrTiO$_3$ (001) substrates (CrysTec GmbH, Germany) were heated to 900\textdegree C (growth temperature) in the hybrid MBE system (Scienta Omicron, Inc). Growth was preceded by 20 min of oxygen cleaning via 250 W RF oxygen plasma with an oxygen pressure of 5 × 10$^{-6}$ Torr (Mantis, UK). SrTiO$_3$ growth was performed by co-deposition of Strontium and Titanium tetraisopropoxide (TTIP) precursor in presence of oxygen plasma. Strontium (99.99\% Sigma Aldrich), was provided via thermal sublimation from an effusion cell kept at 463.5\textdegree C to achieve a beam equivalent pressure (BEP) of 4 × 10$^{-8}$ Torr. TTIP precursor, 99.999\% Sigma Aldrich, was supplied using a gas inlet system consisting of a linear leak valve and a Baratron monometer \cite{21_jalan_jourmatres}. During growth, oxygen was supplied using the same oxygen plasma parameters that were used for oxygen cleaning. After growth, the sample was annealed in a tube furnace at 900\textdegree C for 1 hour in an oxygen rich environment. Ex-situ surface characterization using atomic force microscopy (AFM) was performed to check the surface morphology and high-resolution X-ray diffraction (HRXRD) was performed using Rigaku SmartLab XE thin film diffractometer equipped with Cu K${\alpha}$ radiation. 

From rocking curve data for the STO (002) peak, the FWHM is 0.00716 deg, compared to 0.009 deg for a representative (001)-oriented STO substrate. The lack or rocking curve peak broadening is evidence of a high quality, stoichiometric MBE-grown film. Reciprocal space mapping (RSM) of similar films can be found in \cite{22_jalan_applphyslet}, showing perfect overlap between substrate and film reflections. Fringes in XRD scans of stoichiometric films are routinely observed and have been investigated in \cite{09_lebeau_applphyslet}. The authors found that the fringes are related to offsets as small as a few picometers between film and substrate, and may be caused by carbon contamination at the film/substrate interface.

\section{Device fabrication}

\begin{figure}[htbp]
\centering
\includegraphics[width=7in]{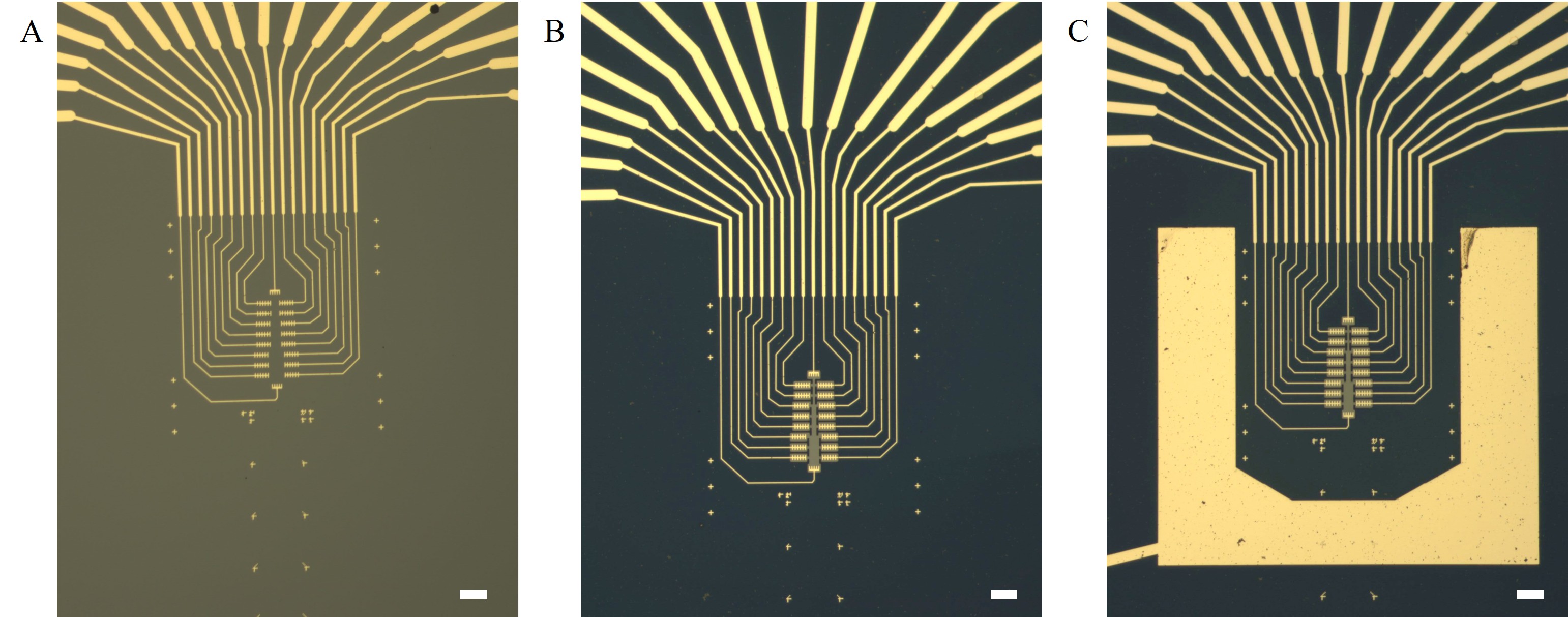}
\caption{\label{SM_figure_device} \textbf{Device fabrication steps.} Optical images of the device taken after lift-off procedure for the (A) ohmic contacts, (B) mesa insulation, (C) gate contact. All scale bars are 50 microns.}
\end{figure}

The ionic liquid gated Hall bar device was patterned in three e-beam lithography steps on poly(methyl methacrylate) (PMMA) 950K 8\% solution in anisole. Figure \ref{SM_figure_device} shows optical images of the device after each step. The first pattern is for ohmic contact lines and bonding. The same resist pattern is exposed to Argon ion milling and e-beam evaporation of 10/80 nm of Ti/Au, followed by lift-off in acetone. The second pattern defines the Hall bar contour by room temperature sputtering deposition of 80-nm thick SiO$_2$ onto the second resist pattern, followed by lift-off in acetone. The third patterning step defined the U-shaped gate and its bonding pad by e-beam evaporation of 40/120 nm Ti/Au, followed by lift-off in acetone.

\clearpage
\section{Preliminary transport characterization above 1.6 K}

Preliminary transport characterization of the device was performed in a pumped liquid He$^4$ cryostat with a variable temperature insert and a superconducting 14 Tesla magnet. Complete and rapid characterization of transport between all voltage probes on the Hall bar was enabled by the home-built multi-terminal lock-in setup described in \cite{23_mikheev_revsciinst}.

\begin{figure}[htbp]
\centering
\includegraphics[width=5in]{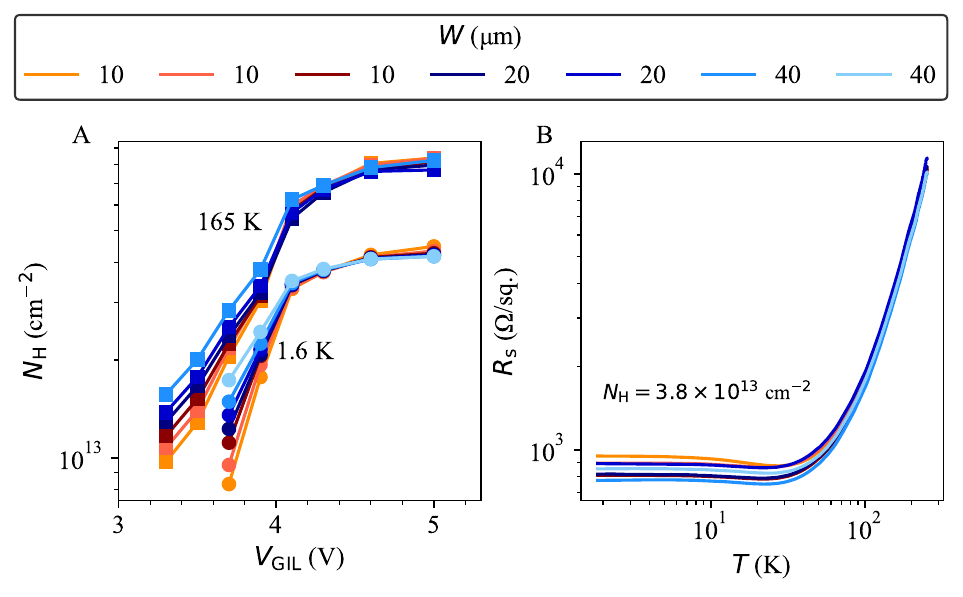}
\caption{\label{SM_figure3} \textbf{Tuning carrier density and Hall Effect measurements.} (A) Carrier density as a function of $V_{\mathrm{GIL}}$. Hall Effect measurements are shown at 165 K and at a base temperature of 1.6 K. (B) Sheet resistance during the cooldown with $N_{\mathrm{H}} = 3.8\times10^{13} \mathrm{cm}^{-2}$. The different traces correspond to four terminal measurement with adjacent voltage probes along the Hall bar channel in Fig. \ref{SM_figure_device}C, with the local channel width indicated in the legend.}
\end{figure}

Eight cooldowns were performed with ionic liquid gate voltage setpoints between 3.3 and 5 V, producing gradual tuning of 2D electron density at the surface of the SrTiO$_3$ film. Hall densities measured along the Hall bar at 165 K and and 1.6 K are shown in Figure \ref{SM_figure3}A. The approximately two-fold reduction can be attributed to carrier freeze-out, as commonly reported in SrTiO$_3$ heterostructure 2DEGs \cite{11_liu_physrevlett, 13_liu_physrevx, 13_shi_applphyslett} or to a change in Hall scattering factor \cite{22_jalan_sciadv}. Notably, this Hall density reduction with temperature was not observed in ionic liquid devices on SrTiO$_3$ single crystal surfaces \cite{21_mikheev_sciadv,23_mikheev_natelec}.

Substantial non-uniformity in 2DEG density accumulation along the Hall bar channel is observed at lowest measure Hall densities (1-2 $\times$  10$^{13}$ cm$^{-2}$, limited by ohmic contact resistance). Figure \ref{SM_figure3}B shows uniform metallic transport across the entire Hall bar channel for a representative cooldown at higher Hall density (3.8 $\times$  10$^{13}$ cm$^{-2}$ at 1.6 K).

\begin{figure}[htbp]
\centering
\includegraphics[width=7in]{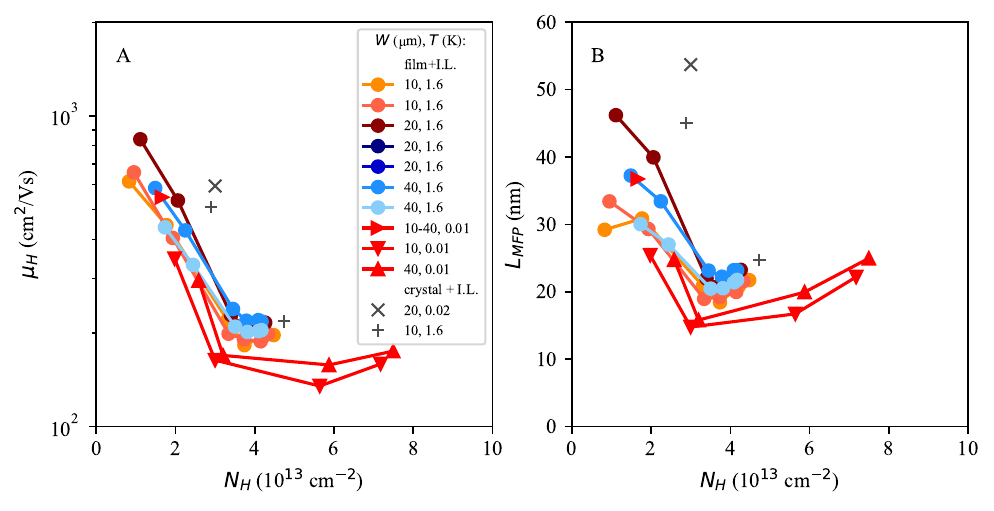}
\caption{\label{figSM_mu_lmfp} \textbf{Low temperature disorder metrics.} (A) Hall mobility and (B) mean free path. ''Film+I.L.'' and ''crystal+I.L.'' labels distinguish ionic liquid gated devices on SrTiO$_3$ film and single crystal surfaces, respectively. Channel width between voltage probes and measurement temperature are indicated in the legend}
\end{figure}

Figure \ref{figSM_mu_lmfp}A summarizes mobility values measured at low temperature, including both the preliminary characterization at 1.6 K for the entire device channel and the subsequent dilution refrigerator cooldowns monitoring select voltage probes on the channel. The $W =$ 10-40 $\mu$m point indicates measurement between voltage probes on the opposite ends of the Hall bar channel. A clear trend is observed for Hall mobility decreasing with Hall density from 600-900 cm$^2$/Vs near  $1\times10^{13} \mathrm{cm}^{-2}$ to 150-200 cm$^2$/Vs near $3\times10^{13} \mathrm{cm}^{-2}$. The gray cross markers indicate low-temperature mobility for similar devices on SrTiO$_3$ single crystal surfaces \cite{21_mikheev_sciadv, 23_mikheev_natelec}, showing a slightly higher mobility and a similar trend with Hall density. Hall mobility was converted into electronic mean free path in Figure \ref{figSM_mu_lmfp}B, using $L_\text{MFP}=\mu_\text{H} e^{-1}\hbar \sqrt{2\pi N_\text{H}}$. The observed $L_\text{MFP}=$ 10-50 nm is within the typically observed range for SrTiO$_3$ 2DEG devices \cite{23_mikheev_natelec}.

\begin{figure}[htbp]
\centering
\includegraphics[width=7in]{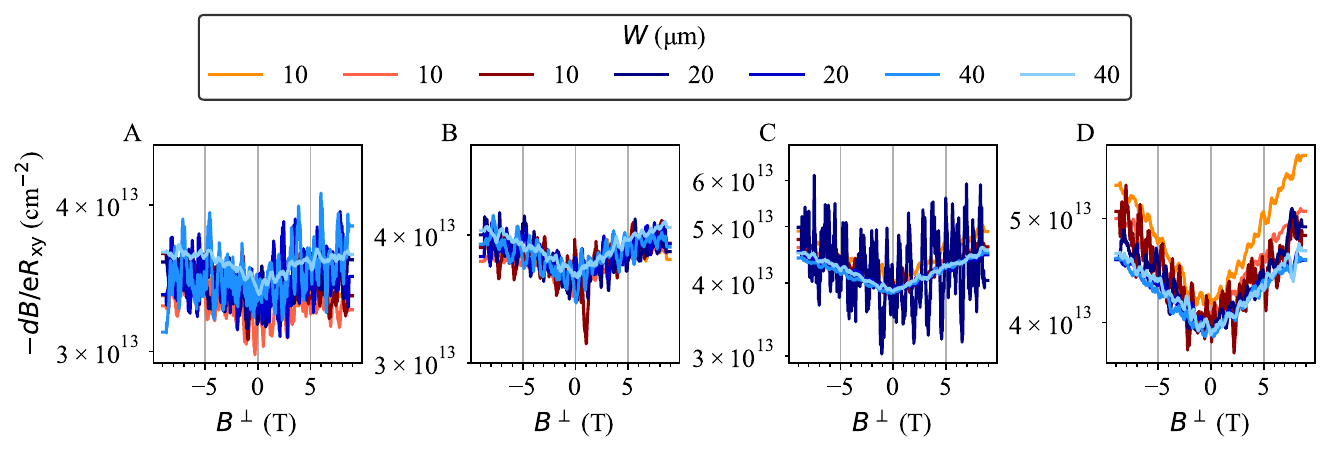}
\caption{\label{SM_figure4} \textbf{High Field Hall Effect measurements for various cooldowns.} (A), (B), (C), (D) Carrier density using a derivative of Hall resistance $R_{\mathrm{xy}}$ w.r.t magnetic field for various cooldowns.}
\end{figure}

Preliminary Hall Effect measurements at 1.6 K were performed up to $\pm$ 9 T, presented as a derivative of $R_{\mathrm{xy}}$ w.r.t magnetic field. The observed non-linearity is typically below 10\% and only reaches approximately 20\% at high Hall density. Neglecting this non-linearity, low-field values of the Hall coefficient was used for Hall density extraction in this work.

\clearpage
\section{Ionic Liquid Gate Voltage Tuning in the Dilution Refrigerator System}
\subsection{Room Temperature Tuning}

This section details the procedure for ramping up the ionic liquid gate voltage, $V_{\mathrm{GIL}}$), at room temperature and the accumulation of carriers in the Hall Bar channel. The sample was mounted onto a Qdevil QBoard sample holder, and electrically contacted using Al wirebonds.  A small droplet of DEME-TFSI ionic liquid was deposited on the gate and the device. The sample mount was then installed into a bottom-loading fast sample exchange system of a Bluefors LD400 dilution refrigerator. To minimize moisture contamination in the ionic liquid, the sample was pumped overnight at room temperature before measurements.

\begin{figure}[htbp]
\centering
\includegraphics[width=3.5in]{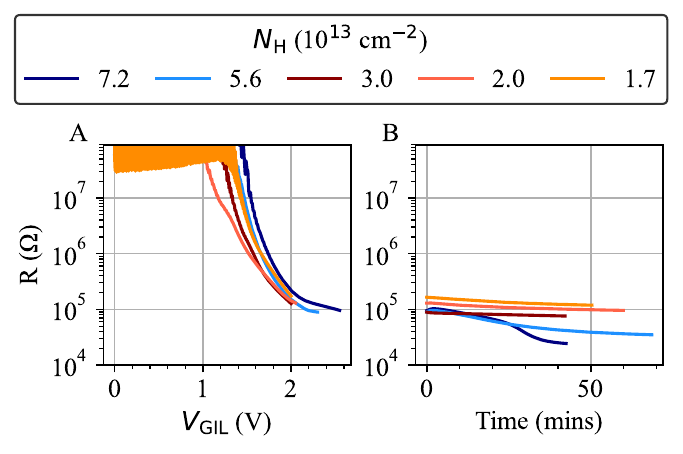}
\caption{\label{SM_figure1} \textbf{2DEG channel tuning with ionic liquid gate voltage at room temperature.} (A) $V_{\mathrm{GIL}}$ was ramped while monitoring 2-terminal resistance through the channel. (B) Cooldowns were started after the device state stabilized. The legend indicates the subsequently measured Hall density at low temperature.}
\end{figure}

The device channel was defined by using insulating SiO$_2$ to separate the area around the channel from the ionic liquid. $V_{\mathrm{GIL}}$ was ramped at room temperature, polarizing the liquid and accumulating a 2DEG on the exposed surface of the undoped STO thin film. A small DC source voltage, $V_{\mathrm{S}}$ = 1 mV, was applied on one of the contacts while all others were kept grounded. This allowed monitoring of the 2-terminal resistance of the channel while ramping the gate voltage. A gradual transition from an insulator to a conductor started above $V_{\mathrm{GIL}}$ = 1 V as shown in Figure \ref{SM_figure1}A. The sample was then maintained at this constant $V_{\mathrm{GIL}}$ to allow some time for stabilization as shown in Figure \ref{SM_figure1}B and during the cooldown process. The ionic liquid freezes near 200 K, fixing the 2DEG charge until the sample is thermally cycled to room temperature to adjust $V_{\mathrm{GIL}}$.

For milliKelvin measurements, The low frequency measurement circuit of the BlueFors LD400 dilution refrigerator included Qdevil Qfilter low pass filters at the mixing chamber stage and low pass RC filters at the sample holder. AC voltage excitation (with frequency ranging from 15 to 30 Hz) was sourced into the device channel through a large bias resistor to limit device current to 10 nA or less. Three SR860 lock-ins were used to monitor drain current and two pairs of voltage probes, amplified by Basel Precision Instruments 1x SP983c-IF IV converter and 2x SP1004 differential voltage amplifiers. During all dilution refrigerator cooldowns, we monitored the voltage across the leads on the 10 and 40 $\mu$m wide channels, except in the cooldown with the lowest carrier density, voltage across the entire hall bar was monitored due to issues with the remaining Ohmic contacts. 

\subsection{Base Temperature Tuning}
\label{base_temp_tuning}

\begin{figure}[htbp]
\centering
\includegraphics[width=7in]{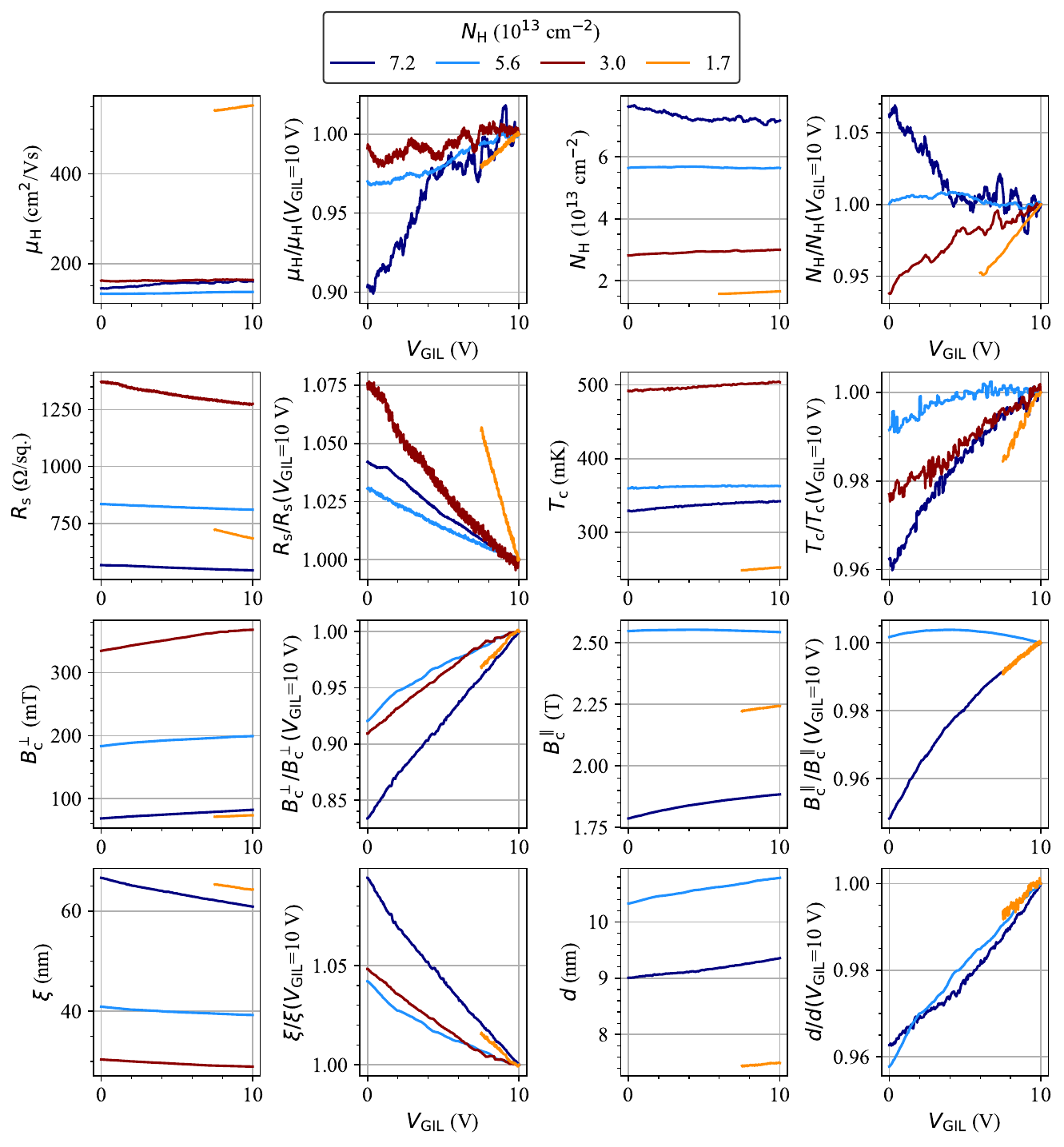}
\caption{\label{SM_figure2} \textbf{Tuning various material and superconducting properties with ionic liquid gate voltage.} First and third columns: Raw measured values with $V_{\mathrm{GIL}}$. Second and fourth columns: Values normalized to $V_{\mathrm{GIL}}$ = 10 V.} 
\end{figure}

After the sample reached base temperature and the ionic liquid is frozen, changing $V_{\mathrm{GIL}}$ provided marginal electrostatic tuning of the  2DEG via electric field lines through the SrTiO$_3$ substrate \cite{23_mikheev_natelec}. All measurements at base temperature were performed with $V_{\mathrm{GIL}}$ = 10 V unless otherwise specified. The effect of $V_{\mathrm{GIL}}$ tuning on various quantities such as carrier density $N_{\mathrm{H}}$ and electron mobility $\mu_{\mathrm{H}}$, normal state sheet resistance $R_{\mathrm{n}}$, critical temperature $T_{\mathrm{C}}$, critical magnetic field $B_{\mathrm{c}}$, superconducting coherence length $\xi$ and 2DEG thickness $d$ are shown in Figure \ref{SM_figure2}.

The effect of $V_{\mathrm{GIL}}$ tuning on $R_{\mathrm{s}}$ is as expected - $R_{\mathrm{s}}$ decreases with gate voltage. This would indicate  an increase in $\mu_{\mathrm{H}}$ and/or $N_{\mathrm{H}}$. To measure this, we performed Hall effect measurements as a function of $V_{\mathrm{GIL}}$, and extracted $N_{\mathrm{H}}$. The electron mobility was calculated using the normal state sheet resistance and Hall density. For cooldowns with low initial Hall density, electrostatic tuning by $V_{\mathrm{GIL}}$ is consistent with a small capacitive gate effect, as expected for a gate contact 200 $\mu$m away from the channel. At higher initial Hall densities, $V_{\mathrm{GIL}}$ does not have a measurable effect on Hall density and primarily tunes Hall mobility $\mu_{\mathrm{H}}$

The effect of $V_{\mathrm{GIL}}$ on $T_{\mathrm{c}}$ was consistently positive, even in the "overdoped" regime of Hall density, where a capacitive gate modulation of Hall density would be expected to produce the opposite effect. The positive and non-monotonic gate effects on perpendicular and in-plane critical fields results from a competition between tuning of the two relevant length scales:  $V_{\mathrm{GIL}}$ decreases the coherence length $\xi$ (as expected from Equation 4 in the main text), but increases 2DEG thickness $d$. The  2DEG confinement tuning by $V_{\mathrm{GIL}}$ is also consistent with the trend in Hall mobility. Pushing a larger proportion of the electrons deeper into the crystal (i.e. increasing $d$) is expected to decrease surface scattering (i.e. increasing $\mu_{\mathrm{H}}$). It is also notable that the normalized modulation of 2DEG confinement $d/d(V_{\mathrm{GIL}} = \mathrm{10 V})$ appears to collapse on a single curve for several cooldowns across a broad range of Hall density, suggesting universal (or nearly so) electrostatics of gate effect screening.

\section{Superconducting Transitions with Magnetic Field}

\begin{figure}[h]
\centering
\includegraphics[width=7in]{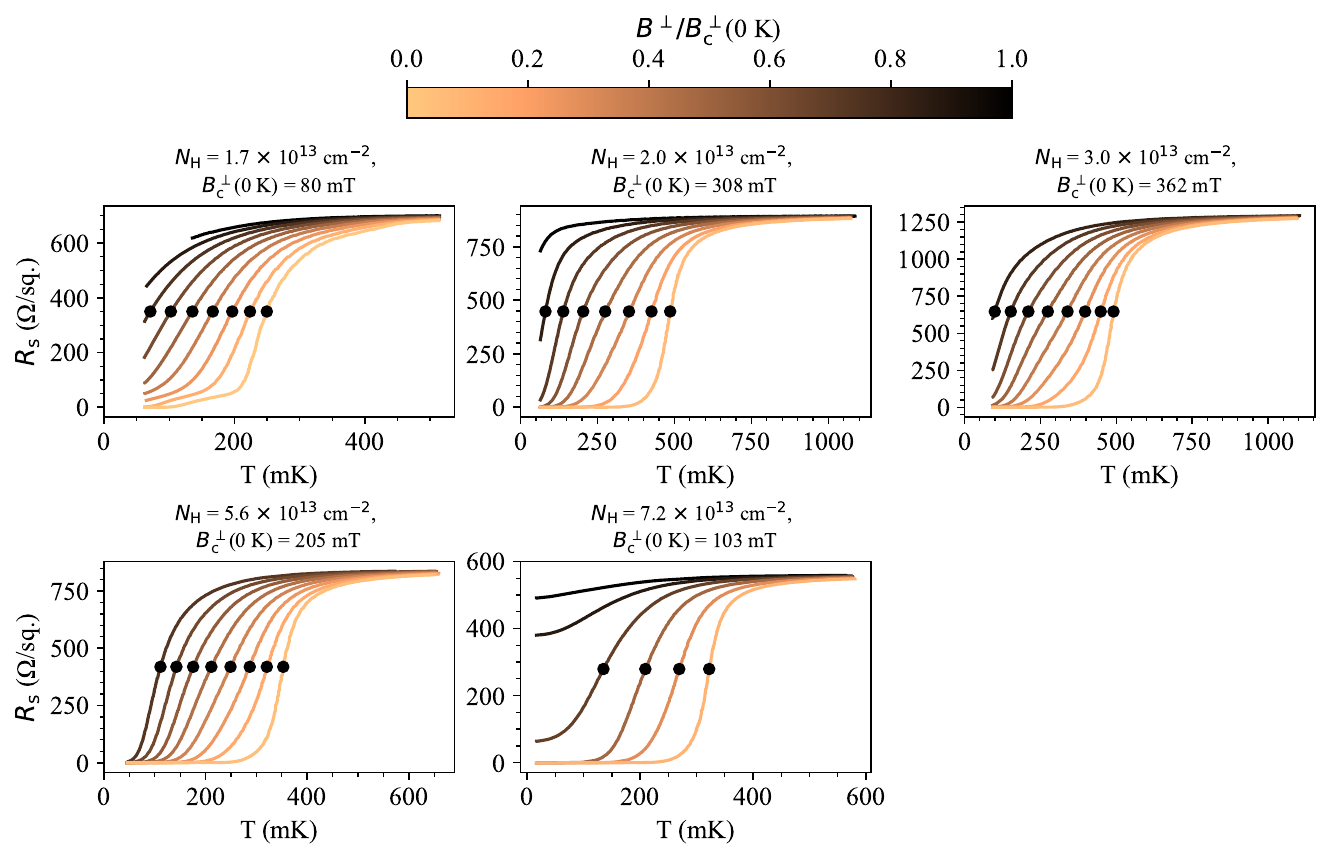}
\caption{\label{SM_figure8} \textbf{Superconducting transitions in perpendicular magnetic field.} Temperature dependence of the 2DEG sheet resistance in constant perpendicular magnetic filed. Only selected traces are shown for clarity. Black circles indicate the identified transition point for each trace. The field scale is normalized to the zero temperature critical field extracted from Ginzburg-Landau fits.}
\end{figure}
\begin{figure}[h]
\centering
\includegraphics[width=7in]{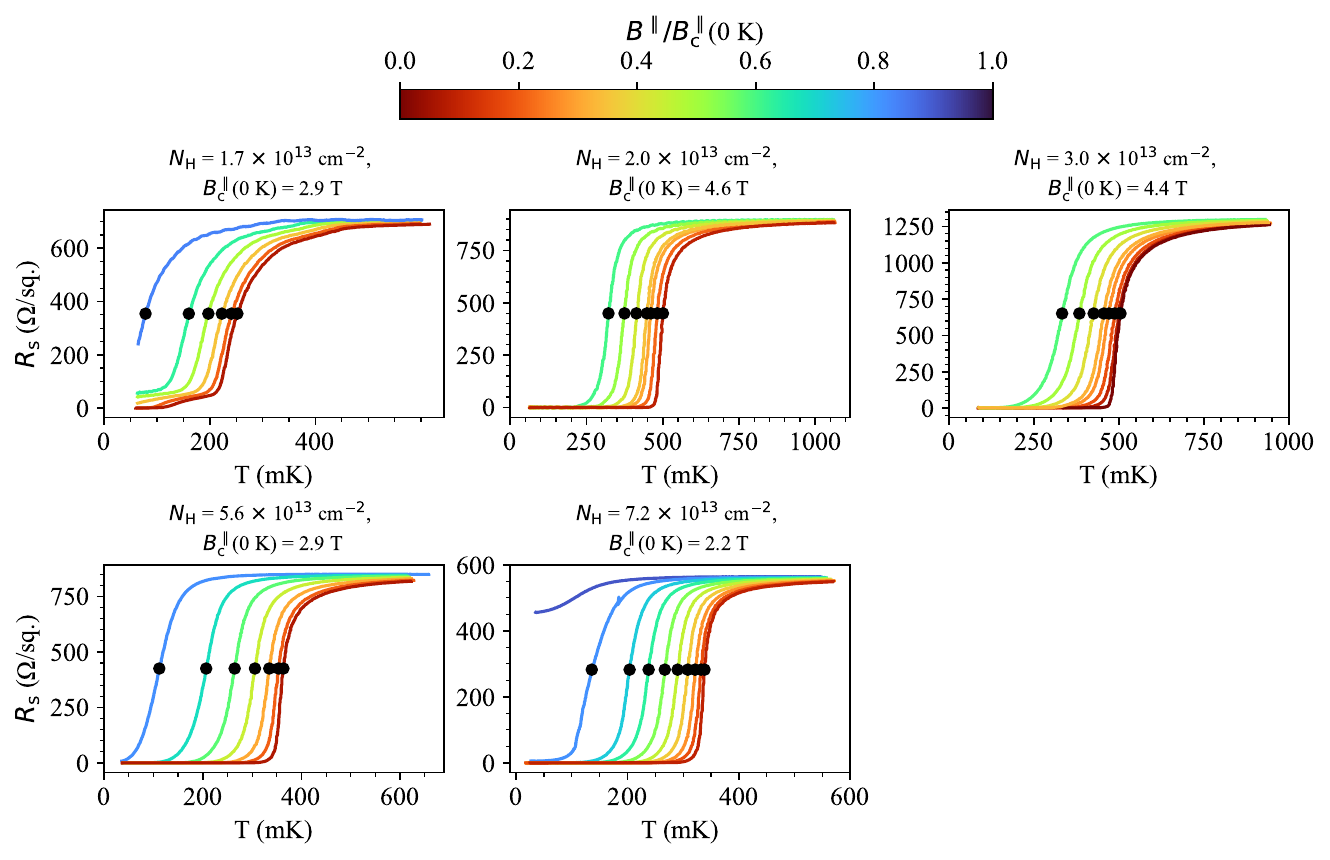}
\caption{\label{SM_figure9} \textbf{Superconducting transitions in in-plane magnetic field.} Same as Figure \ref{SM_figure8}, but for in-plane magnetic field orientation. }
\end{figure}

This section presents the raw data for superconducting transitions in constant magnetic field (Figures \ref{SM_figure8} and \ref{SM_figure9}). This data was used for Ginzburg-Landau fitting to extract 0 K values of critical fields, the superconducting coherence length, and the superconducting layer thickness.

\clearpage

\section{Superconducting Transitions and Effective Medium Theory}
\label{suppinfo_emt}

This supplemental section presents effective medium theory (EMT) model for superconducting transition broadening as an alternative to the superconducting fluctuation model described in the main text. A highly inhomogeneous superfluid density was observed from scanning probe measurements of LAO/STO 2DEGS, with superconducting puddles in the microscale \cite{11_hwang_natphys, 16_moler_physrevb}. Within EMT, superconducting puddles connected by normal-state regions are considered as a network of random resistors with different local resistances and a distribution of critical temperatures. Such a system can then be modeled to approximate the resistance of the entire network \cite{11_caprara_physrevb, 19_caprara_physrevb}. This approach can accommodate probability distributions for the $T_{\mathrm{c}}$'s with characteristic values $T_{\mathrm{n}}$, widths $\sigma_{\mathrm{n}}$ and weights $\omega_{\mathrm{n}}$ for 'N' modes. The network resistance $R_{\mathrm{em}}$  is:

\begin{equation}
\label{eqres_emt}
\frac{R_{\mathrm{em}}(T)}{R_{\mathrm{normal}}} = \sum_{\mathrm{n}=1}^{\mathrm{N}} \omega_{\mathrm{n}} \left[ \mathrm{erf} \left( \frac{T - T_n}{\sqrt{2}\sigma_n} \right) - \mathrm{erf} \left( \frac{T_{\mathrm{c}}^{\mathrm{\alpha}} - T_n}{\sqrt{2}\sigma_n} \right) \right].
\end{equation}
The weights $\omega_{\mathrm{n}}$ describe the contribution to superconductivity from each mode. So, $\sum_{\mathrm{n}}\omega_{\mathrm{n}}=1$. The percolation threshold temperature $T_{\mathrm{c}}^\alpha$ is fixed by:
\begin{equation}
\label{eqtc_alpha}
\omega_1\mathrm{erf}\left(\frac{T_{\mathrm{c}}^\alpha-T_1}{\sqrt{2}\sigma_1}\right) + \omega_2\mathrm{erf}\left(\frac{T_{\mathrm{c}}^\alpha-T_2}{\sqrt{2}\sigma_2}\right) = 0.
\end{equation}

We used the equations \ref{eqres_emt} and \ref{eqtc_alpha} to fit our transitions (R(T)) to a distribution with one and two modes. The effective medium theory does not require the entire device to be superconducting to reach zero resistance: islands of normal-state and superconducting 2DEG can co-exist as long as the superconducting puddles percolate through the device.

\begin{figure}[htbp]
\centering
\includegraphics[width=7in]{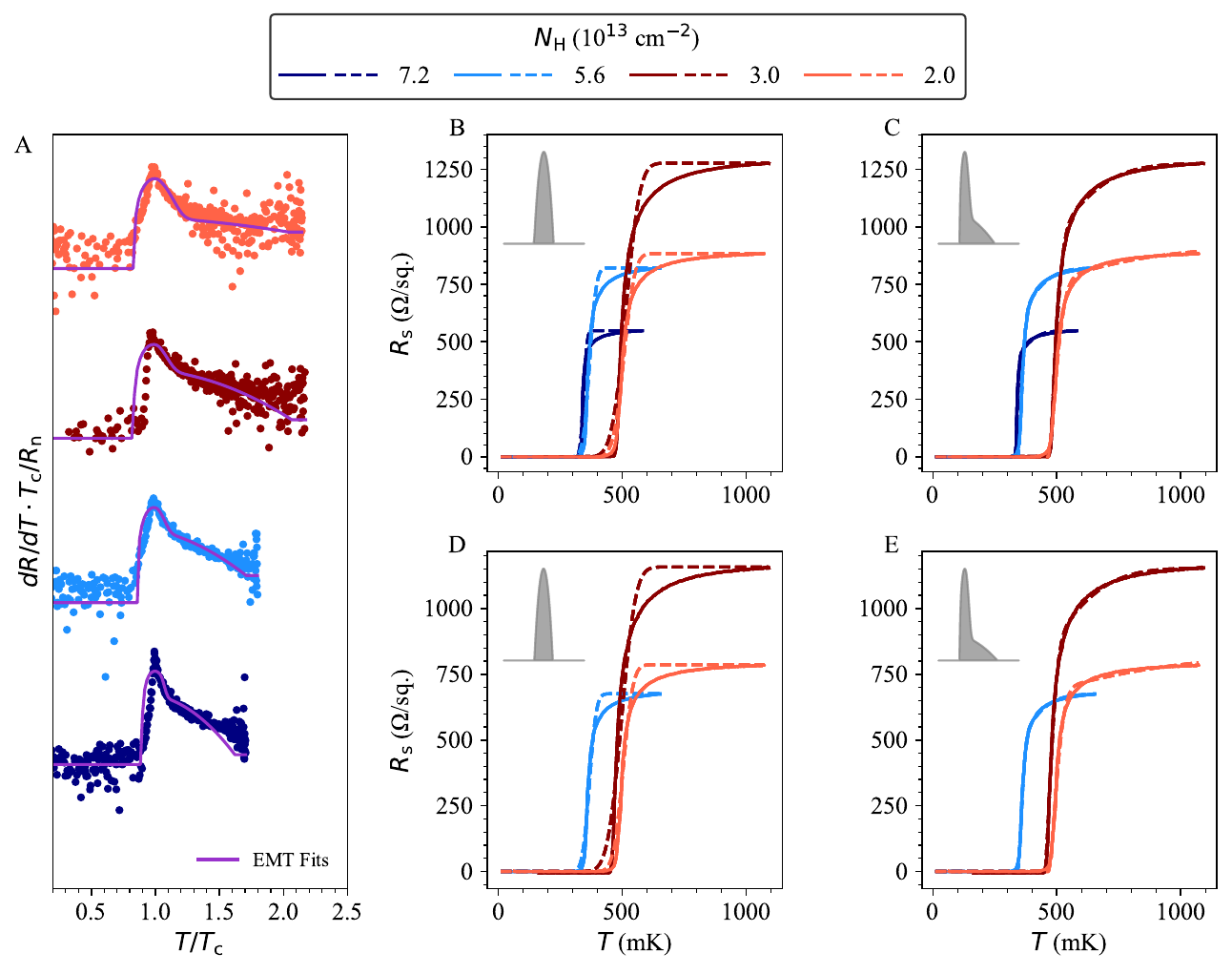}
\caption{\label{SM_figure5} \textbf{Fits to effective medium theory} (A) Superconducting transitions plotted as normalized temperature derivatives of resistance for the 10 $\mu$m channel. The y-axis is in log scale, with arbitrary units. The fits are to a bimodal distribution of $T_{\mathrm{c}}$. (B) \& (D) Fits to a unimodal distribution for the 10 and 40 $\mu$m channels respectively. (C) \& (E) Fits to a bimodal distribution for the 10 and 40 $\mu$m channels respectively. The insets for (B) - (E) show the distribution profile in log scale.}
\end{figure}

Figure \ref{SM_figure5} shows these fits to effective medium theory. Equation \ref{eqres_emt} was used, with the free parameters being $\omega_{\mathrm{n}}$, $T_{\mathrm{n}}$, and $\sigma_{\mathrm{n}}$. $T_{\mathrm{c}}^{\alpha}$ is the critical temperature where the percolation threshold is reached. The transition at the lowest carrier density was excluded since the measurement was across the entire Hall Bar, and the transition at the highest carrier density for the 40 $\mu$m channel has also been excluded. Both of these transitions have a distinct 'two-step' nature. This is likely due to inhomogeneity that arises from disorder due to low carrier density or tetragonal domain walls. A distribution with 3 modes is necessary to fit such transitions. For all cooldowns, a single mode fit (shown in Figures \ref{SM_figure5}B and \ref{SM_figure5}D) fails to capture the asymmetry between a narrow tail below $T_{\mathrm{c}}$ and a wide head above $T_{\mathrm{c}}$.

\begin{figure}[htbp]
\centering
\includegraphics[width=7in]{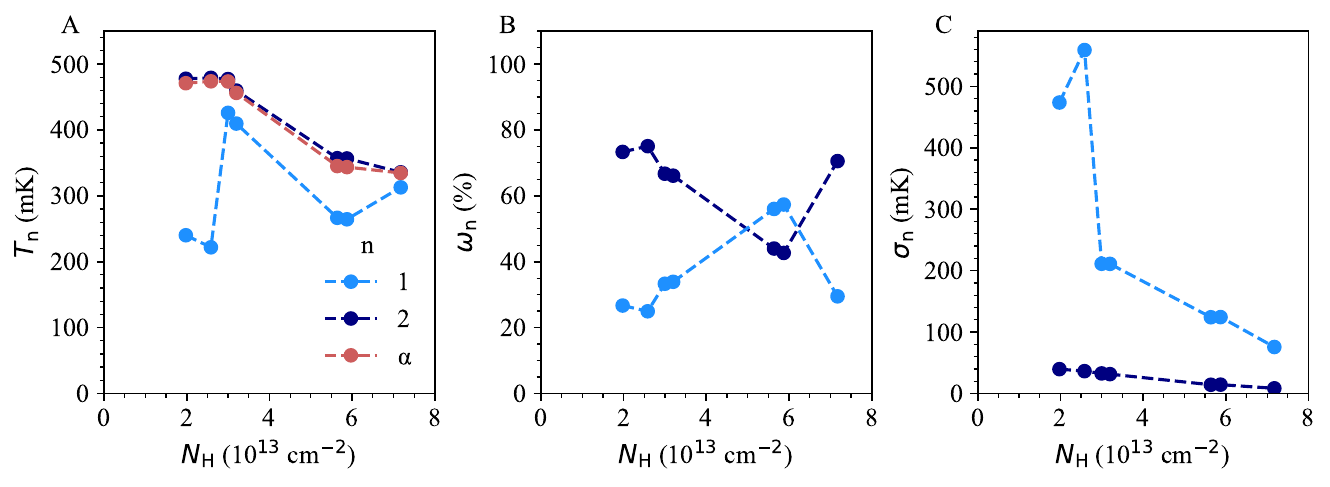}
\caption{\label{SM_figure6} \textbf{Results from bimodal fits to effective medium theory} (A), (B), (C) Extracted values for critical temperature, weight, and width for each mode from fits.}
\end{figure}

The extracted fit parameters are shown in Figure \ref{SM_figure6}. There are two modes in the fit - the first mode ($n=1$) has a lower $T_{\mathrm{c}}$ but it is wider. This mode accounts for the long onset of superconductivity above $T_{\mathrm{c}}$, as well as some of the 'tail' below $T_{\mathrm{c}}$. The second mode ($n=2$) has a higher $T_{\mathrm{c}}$ and it is narrower. This mode accounts for the sharp transition with a large resistance drop near the global network $T_{\mathrm{c}}$ at $R_{\mathrm{n}}/2$. Because of the percolation threshold, the exact $T_{\mathrm{c}}$ and width for the first mode are only weakly determined by the fit. Comparable fit traces can be obtained with any $T_{\mathrm{c}}$ below $T_{\mathrm{c}}^{\alpha}$ or even $T_{\mathrm{c}}=0$. The latter case represents a minority of normal state islands embedded in a fully percolated superconducting matrix with zero total network resistance. Previous examination of $dR/dT$ near $T_{\mathrm{c}}$ in SrTiO$_3$ heterostructures \cite{18_hwang_ncomms} identified intermediate transitions from decoupled islands with higher $T_{\mathrm{c}}$ to a globally coherent superconducting state within both the islands and the surrounding matrix with lower $T_{\mathrm{c}}$. This is a meaningful precedent for the high and low $T_{\mathrm{c}}$ 2DEG fractions represented by the $n=$ 1 and 2 modes, respectively. However, the wide $\sigma$ up to 500 mK of the low $T_{\mathrm{c}}$ implies a minority fraction of the 2DEG with $T_{\mathrm{c}}$ around 900 mK. Previous reports of inhomogeneity only detected superconducting patches below 400 mK \cite{11_caprara_physrevb, 16_moler_physrevb, 18_hwang_ncomms, 19_caprara_physrevb}.

A related critical issue with this EMT description is in reconciliating it with the competing superconducting fluctuation model presented in Figure 3. The collapse of data from several cooldowns onto a single $\Delta\sigma(\varepsilon)$ curve following from AL and MT fluctuation theories motivated our choice of centering that model in the main text for explaining  transition broadening above $T_{\mathrm{c}}$. The paraconductivity scaling from AL and MT force a very rapid resistance decrease to zero below $T_{\mathrm{c}}$. So the fluctuation model must be limited to temperatures above $T_{\mathrm{c}}$ and cannot describe the resistance tail below $T_{\mathrm{c}}$. The EMT model is capable of describing the entire temperature range, but without a compelling justification for the high $T_{\mathrm{c}}$ fraction of the 2DEG.

Speculatively, we suspect that an accurate full treatment of this resistive transition needs to account for a gradual shift from superconducting fluctuations dominating above $T_{\mathrm{c}}$ to an effective medium description below $T_{\mathrm{c}}$. The authors are not aware of a full theoretical treatment of this problem in this literature beyond a simplified case of a single broadened mode with AL fluctuations only in \cite{11_caprara_physrevb}.

\clearpage

\clearpage
\section{AL-MT Fluctuations Extended}

\begin{figure}[htbp]
\centering
\includegraphics[width=6in]{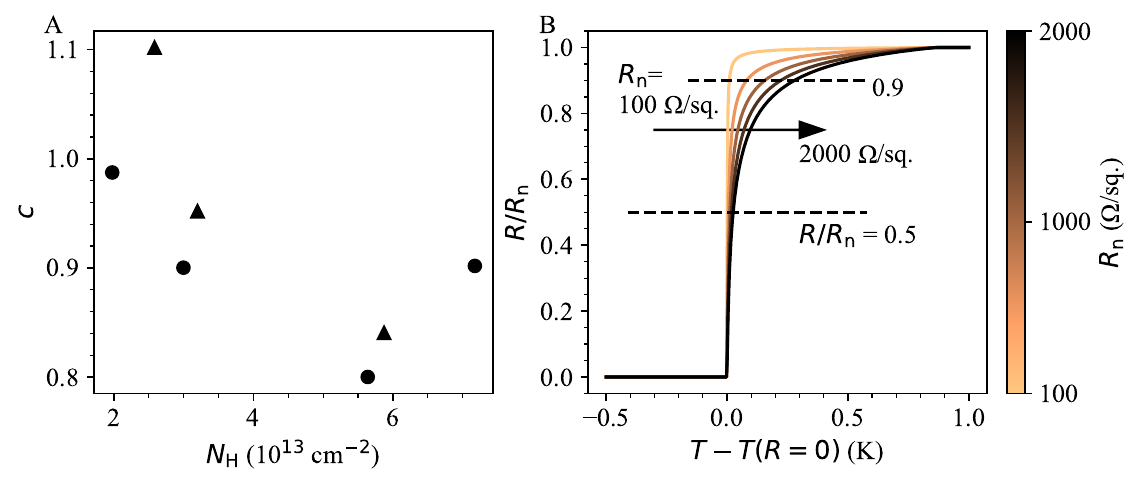}
\caption{\label{SM_figure7} \textbf{Results from AL-MT Fits and Predicted Transition Width} (A) The extracted fit values for the total energy cutoff are shown across the superconducting dome. The $\bullet$ and $\blacktriangle$ show the values for the 10 and 40 $\mu$m channel widths respectively (B) Expected transitions from AL-MT fluctuations at $\delta=0.22$ and $c=1$ while varying only $R_{\mathrm{n}}$.}
\end{figure}

Figure \ref{SM_figure7}A shows the extracted fit values for the total energy cutoff across the superconducting dome. The slight decrease in $c$ with carrier density is unlikely to be meaningful. At low carrier densities $T_{\mathrm{c}}$ is higher, so the measured temperature range was extended upwards, influencing the fitting of $c$.

Figure \ref{SM_figure7}B shows the expected transition lineshapes from AL-MT fluctuations while keeping the parameters constant and only changing $R_{\mathrm{n}}$. It illustrates that the transition width increases with $R_{\mathrm{n}}$. We extracted $\Delta T=T(R=0.9*R_{\mathrm{n}})-T(R=0.5*R_{\mathrm{n}})$ and plotted it in the inset in Figure 3A as the expected trend in width with $R_{\mathrm{n}}$.

\begin{figure}[htbp]
\centering
\includegraphics[width=7in]{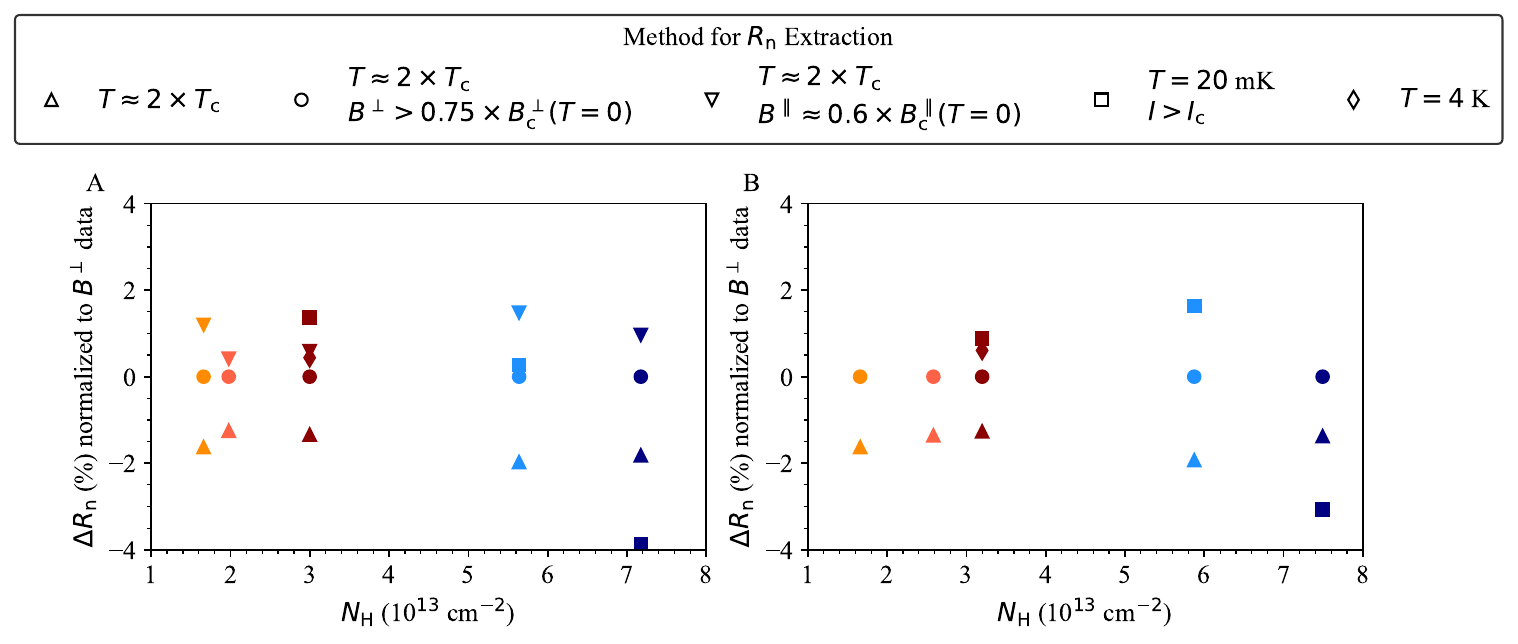}
\caption{\label{fig_rn_comparison} \textbf{Different Methods for $R_{\mathrm{n}}$ determination.} (A), (B) Data for the 10 and 40 $\mu$m channels respectively. Values are shown as percentage deviation from $R_{\mathrm{n}}$ extracted in $B^{\perp}$ above $T_\mathrm{c}$.}
\end{figure}

\begin{figure}[htbp]
\centering
\includegraphics[width=7in]{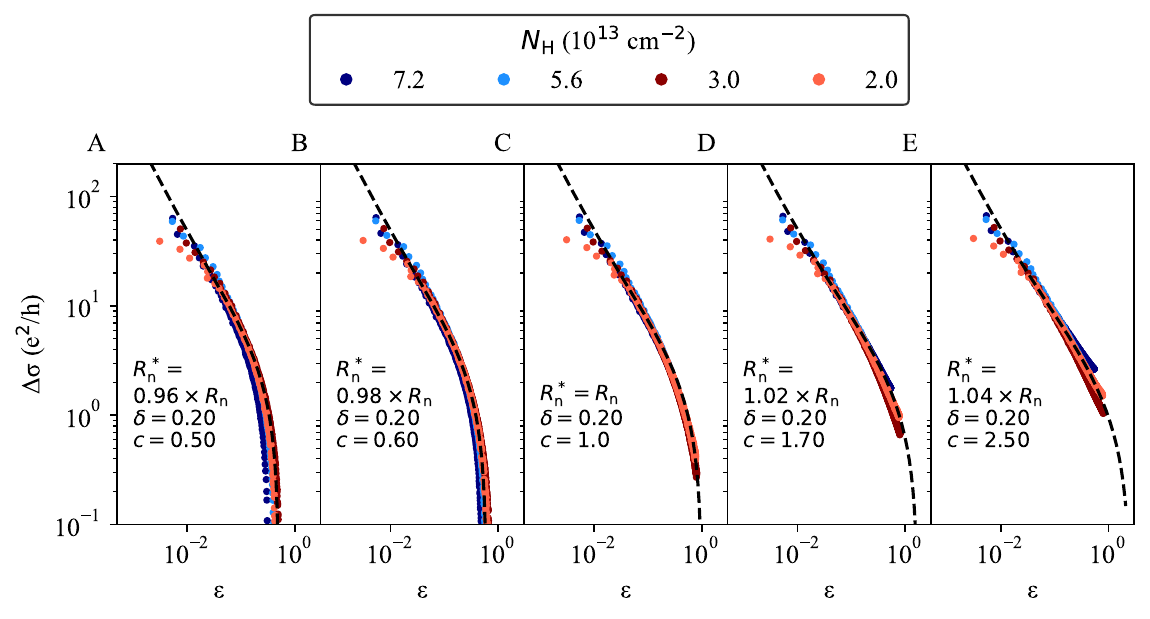}
\caption{\label{fig_almt_comparison} \textbf{Robustness of paraconductivity to uncertainty in $R_{\mathrm{n}}$.} (A)-(E) Paraconductivity above $T_{\mathrm{c}}$ and AL-MT fits with $R_{\mathrm{n}}$ artificially shifted by up $\pm 4\%$s . Panel (c) is same as Fig. 3B in the main text. The black dashed line is a representative fit using the combined AL-MT model using the same $\delta$ and different cutoff ($c$) parameters.}
\end{figure}

Figures \ref{fig_rn_comparison} and \ref{fig_almt_comparison} address robustness of paraconductivtiy analysis and AL-MT fitting to possible errors in determination of normal state resistance. We explored several ways of suppressing both superconductivity and paraconductivity to access the normal state with combinations of temperature, in- and out-of-plane magnetic field, and DC current. Figure \ref{fig_rn_comparison} shows the percentage spread of extracted $R_\mathrm{n}$ values with respect to our chosen method (above $T_c$ in moderate $B^{\perp}$).This spread is typically below $\pm2\%$, with a few outliers below $\pm4\%$ (likely due to Andreev reflections above the critical DC current).

Figure \ref{fig_almt_comparison} replots paraconductivity data from Figure 3B using artificially adjusted $R_{\mathrm{n}}^*$ within the observed $(1\pm0.04)R_{\mathrm{n}}$ range. This significantly affects the evolution of $\Delta \sigma$ at high temperature, $\varepsilon>0.1$. The slope in the unaffected region below $\varepsilon \approx 0.1$ encodes the physically meaningful information on the superconducting transition shift $\delta$ in the MT mechanism. The uncertainty in determination of $R_\mathrm{n}$ is encoded in the energy cutoff parameter $c$, which governs the fit in the high $\varepsilon$ limit. We do not draw any conclusions on physical mechanisms from the extracted $c$ values.

\clearpage

\section{References}

\renewcommand{\bibsection}{\section*{}}
\bibliography{references}

\clearpage




\end{document}